\documentclass[aps,pra,reprint,groupedaddress,amsmath]{revtex4-1}

\usepackage{graphicx}
\usepackage{dcolumn}
\usepackage{amsmath}
\usepackage{lipsum}
\usepackage{xcolor}
\usepackage{physics}

\begin{document}


\title{Zero-velocity atom interferometry using a retroreflected frequency chirped laser}


\author{I. Perrin}
\affiliation{DPHY,ONERA, Universit\'{e} Paris Saclay, F-91123 Palaiseau, France}

\author{J. Bernard}
\affiliation{DPHY,ONERA, Universit\'{e} Paris Saclay, F-91123 Palaiseau, France}

\author{M. Cadoret}
\email{malo.cadoret@lecnam.net}
\affiliation{LCM-CNAM, 61 rue du Landy, 93210, La Plaine Saint-Denis, France}

\author{Y. Bidel}
\affiliation{DPHY,ONERA, Universit\'{e} Paris Saclay, F-91123 Palaiseau, France}

\author{N. Zahzam}
\affiliation{DPHY,ONERA, Universit\'{e} Paris Saclay, F-91123 Palaiseau, France}

\author{C. Blanchard}
\affiliation{DPHY,ONERA, Universit\'{e} Paris Saclay, F-91123 Palaiseau, France}

\author{A. Bresson}
\affiliation{DPHY,ONERA, Universit\'{e} Paris Saclay, F-91123 Palaiseau, France}


\date{\today}

\begin{abstract} 
Atom interferometry using stimulated Raman transitions in a retroreflected configuration is the first choice in high precision measurements because it provides low phase noise, high quality Raman wavefront and simple experimental setup. However, it cannot be used for atoms at zero velocity because two pairs of Raman lasers are simultaneously resonant. Here we report a method which allows to lift this degeneracy by using a frequency chirp on the Raman lasers. Using this technique, we realize a Mach-Zehnder atom interferometer hybridized with a force balanced accelerometer which provides horizontal acceleration measurements with a short-term sensitivity of $3.2\times 10^{-5}$ m.s$^{-2}$/$\sqrt{Hz}$. We check at the level of precision of our experiment the absence of bias induced by this method. This technique could be used for multiaxis inertial sensors, tiltmeters or atom interferometry in a microgravity environment.
\end{abstract}

\pacs{}

\maketitle



\section{Introduction}
Since their inception, light-pulse atom interferometers (AIs) have proven to be extremely sensitive gravito-inertial sensors measuring gravity \cite{Peters2001,Gillot2014,Hu2013,Bidel2018,Hauth2013}, gravity gradients \cite{McGuirk2002,Sorrentino2014,Duan2014}, rotations \cite{Gustavson1997,Gauguet2009,Tackmann2012,Dutta2016}, appearing as promising candidates to compete with traditional sensors used for geodesy, geophysics, exploration or inertial navigation \cite{Jekeli2005}. Moreover, they have demonstrated to be an invaluable tool in fundamental physics where they are used for measuring physical constants \cite{Muller2018,Fixler2007,Bouchendira2011,Rosi2014}, testing Einstein equivalence principle \cite{Fray2004,Bonnin2013,Schlippert2014,Tarallo2014,Zhou2015}, searching for dark sector particles \cite{Hamilton2015}, and even proposed for gravitational-wave detection \cite{Dimopoulos2008,Chaibi2016} or for measuring free-fall of anti-matter \cite{Hamilton2014}. The principle of a light-pulse AI relies on the use of recoils from photon-atom interactions to coherently split, deflect and interfere matter-waves. Most light-pulse AIs use stimulated two-photon process (Raman or Bragg transitions) to realize the beamsplitters and mirrors required for the interferometer sequence \cite{Kasevich1991}. In this process, the atom coherently absorbs and then emits a photon from a pair of counterpropagative laser beams with different frequencies, resulting in a net momentum transfer of $\hbar k_{\mathrm{eff}}$ at each interaction, where $k_{\mathrm{eff}}$ is the effective wave vector. To perform Bragg-based or Raman-based AIs at their best level of performance, it is beneficial to adress the two-photon transitions in a retroreflected geometry where a single laser beam with two laser frequencies is retroreflected off a mirror. This allows first to reduce the effect of wavefront distortions which affect the sensor's accuracy \cite{Peters2015}. This is beacause the wavefront distortion of the incoming beam cancels out in a retroreflected configuration and only optical elements behind the atoms have to be considered for wavefront distortions (mirror, wave plate, vacuum window). Secondly, it is an efficient way to implement the $k_{\mathrm{eff}}$ reversal technique \cite{Durfee2006} to eliminate some systematics, as well as to reduce interferometer phase noise as most vibration effects on the laser phases are common to the two lasers and cancel out in the two-photon process, apart from vibrations of the mirror.  However, the use of retroreflection for zero velocity atoms along the Raman beam naturally leads to a double diffraction scheme where two stimulated Raman transitions with opposite momentum transfer $\pm \hbar k_{\mathrm{eff}}$ are simultaneously resonant. This double diffraction scheme has been first implemented using Raman transitions to realize an AI for which the separation between the two arms is $2\hbar k_{\mathrm{eff}}$ both in the case of atoms at rest \cite{Leveque2009}, as well as for nonvanishing initial velocities in the case of a gravimeter \cite{Malossi2010}. For the latter, three laser frequencies were mandatory to account for the changing Doppler shift induced by gravity acceleration, hence leading to a more complex setup. However, for onboard applications, where shot-to-shot acceleration variations leads to uncontrolled velocity variations of the atomic sample, even though close to zero velocity, it becomes challenging to adress this double diffraction scheme with high efficiency as the two transitions become partly degenerated. Moreover, the experimental realization of a double diffraction AI geometry is much more demanding than that of a single diffraction as it requires longer Raman pulse duration, colder atomic source and additional blow away beams to get rid of parasitic interferometers. Moroever, the gain in scale factor obtained by increasing the arm separation in a double diffraction scheme is not of interest when the sensitivity of the interferometer is limited by vibrations unless used in differential mode accelerometer for applications in gradiometry \cite{Carraz2014} or for testing the WEP \cite{Zhou2015}. Therefore, in certain situations, single diffraction may be preferable to double diffraction. In this work, we experimentally demonstrate a technique enabling the use of single diffraction two-photon Raman transitions despite zero Doppler shift in the commonly-used retroreflected geometry. By employing a laser frequency chirp, we lift the degeneracy between the two simultaneous resonance transitions. We then apply our technique to the measurement of the horizontal component of acceleration using a Mach-Zehnder style atom interferometer. We achieve a sensitivity of $3.2\times 10^{-5}$ m.s$^{-2}$/$\sqrt{Hz}$ and show that no bias is induced by this method.

\section{Method}
\begin{figure}[tbph]
\centering
\includegraphics[scale=.34]{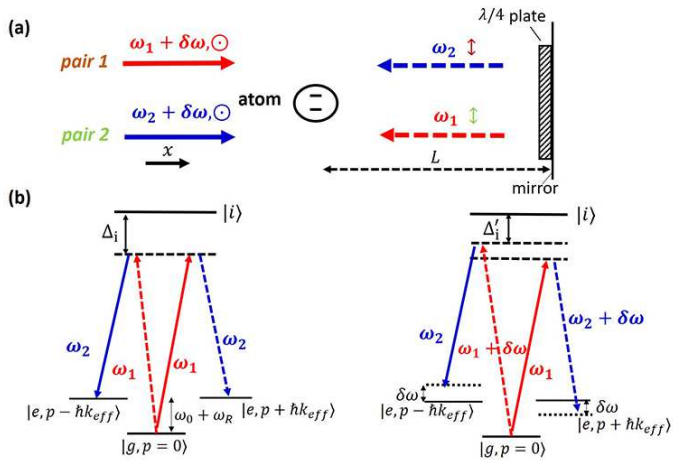}
\caption{(a) Schematic setup of two-photon Raman transitions in the commonly-used retroreflecting geometry. A two-level atom is interacting with two pairs of counterpropagating light fields (pair 1 and pair 2) in lin $\perp$ lin configuration. Applying a frequency chirp $\beta$ on the lasers, changes the incident laser frequency by $\delta \omega=2\pi\beta t_d$. $\Delta_i$ is the one photon detuning. (b) Left: Scheme of the Raman transition between the two hyperfine ground states of an alkaline atom in absence of Doppler shift $(\omega_D=0)$ and without frequency chirp $(\beta = 0)$. Both pairs are simultaneously resonant. Right: Applying a frequency chirp $\beta$ on the Raman laser frequencies, lifts the degeneracy between the two resonant conditions by an amount $2\delta \omega$. After the chirp, the one photon detuning is $\Delta_i^{'}$.}
\label{fig:figure1}
\end{figure}

We present here a general method which can be applied to any two-photon process such as Raman transition or Bragg diffraction. For example, we demonstrate our technique by performing stimulated two-photon Raman transitions between the two hyperfine ground states of $^{87}$Rb (labelled $\ket{g}$ and $\ket{e}$) via an intermediate state (labelled $\ket{i}$) using two lasers of frequencies $\omega_1$ and $\omega_2$ detuned to the red of the $D_2$ line by $\Delta_i$.  The Raman beams are brought to the vacuum chamber via a polarization maintening optical fiber. After passing through the chamber, the laser beams are retroreflected through a quarter-wave plate to rotate the initial polarization into its orthogonal polarization creating two pairs of counterpropagating beams in the horizontal $x$-direction in a lin $\perp $ lin configuration (see Figure \ref{fig:figure1} (a)). Consequently two pairs of beams can drive the two-photon transition between $\ket{F = 1,m_F = 0} \equiv \ket{g} \to \ket{F = 2, m_F = 0} \equiv \ket{e}$. With this polarization configuration, the co-propagating Raman transitions are forbidden.
The detuning $\delta$ from the two-photon resonance is given by:
\begin{equation}
\delta=\omega_1-\omega_2-(\omega_0+\omega_D+\omega_R)
\end{equation}
where $\omega_0$ is the frequency of the hyperfine transition, $\omega_D=\pm\vec{k}_{\mathrm{eff}}.\vec{v}$ is the Doppler shift due to the atomic velocity $\vec{v}$ in the reference frame of the apparatus, and $\omega_R=\frac{\hbar k_{\mathrm{eff}}^2}{2m}$ the recoil frequency shift. Thus, in absence of Doppler shift, both pairs are simultaneously resonant and couple $\ket{g,\vec{p}}\to \ket{e,\vec{p} \pm \hbar \vec{k}_{\mathrm{eff}}}$.
In order to circumvent this problem and lift the degeneracy between the two resonance conditions, we apply a frequency chirp $\beta=\frac{1}{2\pi}\frac{d\omega_1}{dt}=\frac{1}{2\pi}\frac{d\omega_2}{dt}$ on the Raman lasers. As the reflected beams are delayed with respect to the incoming ones by $t_d = \frac{2L}{c}$ (where $L$ is the distance atom-mirror), the incoming laser frequencies will be shifted by $\delta\omega=2\pi\beta t_d$ at the position of the atoms, allowing to detune one transition with respect to the other by $2\delta\omega$. This allows to selectively adress Raman pair 1 or pair 2. This effect can be understood as mimiking an effective atomic velocity in the reference frame of the lasers leading to an equivalent Doppler shift $\omega_D=2\pi\beta t_d$.

\subsection{Experimental setup and lasers}
\begin{figure*}[tbph]
\centering
\includegraphics[scale=.4]{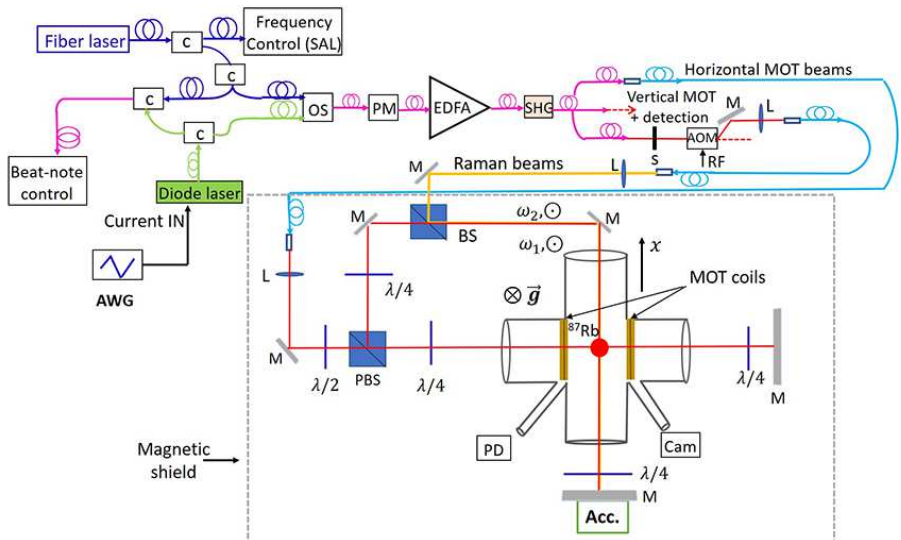}
\caption{Scheme of the laser system and the experimental setup (top view) allowing to perform Raman transitions in the horizontal scheme. SAL, Saturated absorption lock; L, lens; M, mirror; PM, phase modulator; AOM, acousto-optic modulator; S, shutter; PBS, Polarizing beam-splitter; OS, optical switch; BS, beam splitter; AWG, Arbitrary waveform generator; SHG, second harmonic generation; Acc, accelerometer; c, coupler; PD, Photodetector; Cam, camera.}
\label{fig:figure2}
\end{figure*}

The experiment was carried out in the atom interferometer setup described in \cite{Perrin2019}. Atom interferometers usually consist in three-steps: preparation, interferometry and population detection. To perform these functions we use the laser system described in detail in \cite{Theron2017}, based on a frequency doubled fiber bench using two independent lasers sharing the same 5 Watts Erbium-doped fiber amplifier. The laser used to cool and detect the atoms is an erbium DFB fiber laser at 1.5 $\mu$m (output power 20 mW, linewidth 2 kHz) locked relative to the Rubidium transitions using a saturated absorption lock \cite{Theron2015}. The atom interferometry (AI) laser source is a DFB laser diode at 1.5 $\mu$m (Avanex DFB, output power: 10 mW, linewidth 1 MHz). The detuning $\Delta$ of the AI laser from the one photon resonance is controlled using a beat-note between the two lasers at 1.5 $\mu$m . Finally the two lasers are combined at 1.5 $\mu$m by an electro-optical modulator which acts like a continuous optical switch between each laser before seeding the EDFA. The output of the EDFA is sent to the dual-wavelength second harmonic generation bench. In our experiment, the two Raman beam frequencies are generated thanks to a fiber phase modulator \cite{Carraz2012}. The chirp $\beta$ is obtained by directly modulating the input current of the laser diode using a low frequency arbitrary waveform generator (AWG) (Agilent, Model 33250A). We display on Figure \ref{fig:figure2} the laser setup and the optical bench of the experiment.

\subsection{Raman spectroscopy experiment}
\begin{figure}
\centering
\includegraphics[scale=0.45]{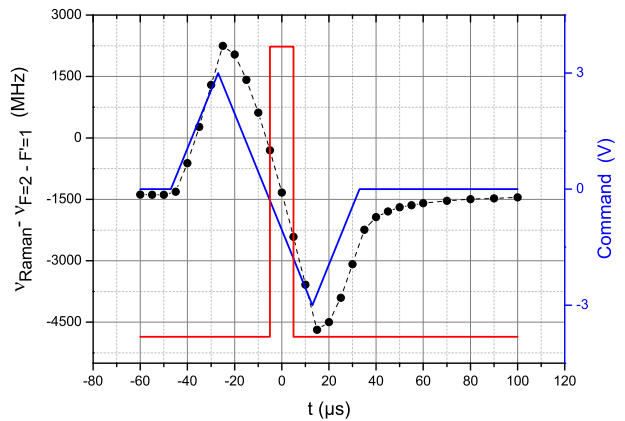}
\caption{Frequency response of the Raman laser $\nu_{\mathrm{Raman}}=\frac{\omega_1}{2\pi}$ to the voltage command applied to the current modulation input. Voltage command (blue line). Measurement of the laser diode frequency response (black dots and line). Raman laser pulse (red line). The measured chirp is $\beta= -210$ MHz.$\mu$s$^{-1}$, leading to a frequency excursion of 2100 MHz during the Raman pulse.}
\label{fig:figure3}
\end{figure}

To investigate our method we first started by implementing Raman spectroscopy. A cold $^{87}$Rb atom sample is produced in a 3 dimensional magneto-optical trap (MOT) loaded from a background vapor pressure in 340 ms. Atoms are further cooled to 2 $\mu$K by means of polarization gradient in 8 ms. The cooling beams are then turned off, and as the atoms freely fall, a microwave $\pi$-pulse followed by a blow-away beam allows to select the atoms in the insensitive ground state $\ket{F = 1,m_F = 0}$ with a horizontal bias magnetic-field of 100 mG. Then, a horizontal Raman laser pulse of duration $\tau = 10\,\mu$s is applied to the atoms 18 ms after their release from the trap. The proportion of atoms in each hyperfine state $F = 1$ and $F = 2$ is then measured using a state selective vertical light-induced fluorescence detection. The cycling time of the experiment is $T_{\mathrm{cycle}} = 500$ ms. In practice, the AWG generates a triangle-wave modulation signal in burst mode, directly applied to the modulation input of the laser diode current controller. The voltage command signal is triggered to the Raman pulse and the chirp duration is fixed to 40 $\mu$s. Consequently, the single-photon frequency excursion is controlled by adjusting the peak-to-peak voltage amplitude denoted $A$. We experimentally measure the frequency response of the laser diode as a function of the voltage amplitude by monitoring the beat-note signal between the laser diode and the fiber laser on a spectrum analyzer (SA). Figure \ref{fig:figure3} displays the frequency response of the laser diode when applying $A = 6$ V peak-to-peak amplitude command. For clarity sake, the frequency response is plotted as the frequency difference beween the Raman laser frequency $\nu_{\mathrm{Raman}}=\omega_1/2\pi$ and the $5S_{1/2},F = 2 \to 5P_{3/2},F' = 1$ transition. In this case, one finds a chirp $\beta=-210$ MHz.$\mu$s$^{-1}$. The delay between the command and the Raman laser pulse is adjusted to ensure a linear frequency response of the laser diode during the Raman pulse.

\begin{figure}[tbph]
\centering
\includegraphics[scale=0.45]{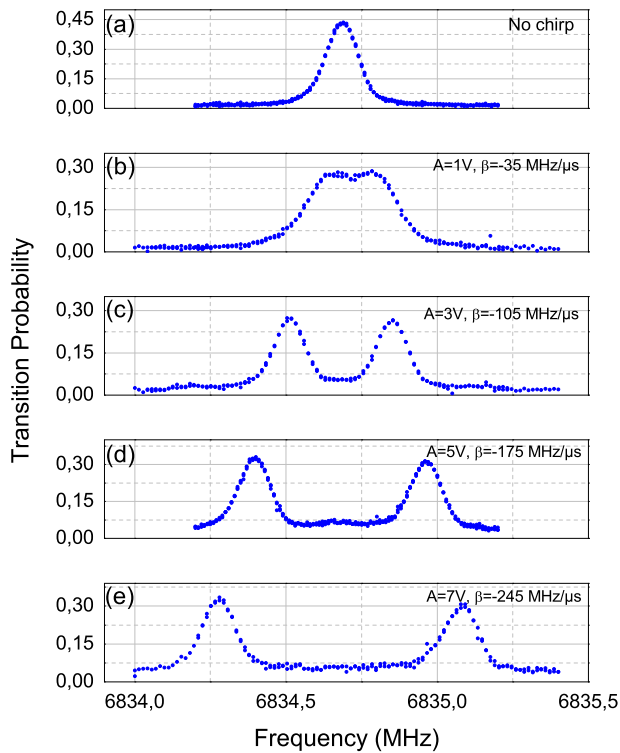}
\caption{ Raman resonance spectra obtained by scanning the Raman frequency difference $(\omega_1-\omega_2)/2\pi$, across the resonance for five different frequency chirps applied on the laser diode. $A$ is the peak-to-peak amplitude voltage of the signal command. (a) For $\beta = 0$ a single peak is observed due to the simultaneous resonant conditions. (b),(c),(d),(e): When applying a frequency chirp two peaks are observed allowing to lift the degeneracy between the two Raman transitions. The two Raman resonances are separated by $2\delta\omega/2\pi$.}
\label{fig:figure4}
\end{figure}

Figure \ref{fig:figure4} diplays the measured transition probability as a function of the Raman frequency difference $(\omega_1-\omega_2)/2\pi$ for different chirp values $\beta$ applied to the laser diode. The Raman laser intensity is adjusted to maximize the transfer efficiency at resonance for a pulse duration of $\tau= 10$ $\mu$s. When no chirp is applied $(\beta = 0)$, only a single peak is observed due to the simultaneous resonant condition. Increasing slightly $\beta$ starts lifting the degeneracy between the two resonant conditions. For $\beta=-35$ MHz.$\mu$s$^{-1}$, the chirp is not important enough to lift the degeneracy between the two transitions. However, for greater values of the chirp two resonance spectra are clearly observed with a frequency separation which increases linearly with the amplitude of the voltage command. Measuring the frequency separation between the two resonance spectra ($2\delta\omega/2\pi$) allows to estimate the atom-mirror distance $L=c\delta\omega/4\pi\beta\simeq 24$ cm in agreement with the expected distance from the trap center to the mirror of the setup.

\section{Atom interferometer}
To further investigate our technique we performed a Mach-Zehnder style AI in a horizontal configuration using a $\pi/2-\pi-\pi/2$ chirped-Raman pulse sequence, with each pulse separated by an interrogation time $T$. With this geometry, the atomic phase shift at the ouput of the interferometer is sensitive to the horizontal acceleration $\vec{a}$ of the atoms relative to the reference mirror. In the limit of short, resonant pulses, the phase shift is then given by: $\Delta\phi= \phi_1 -2\phi_2 + \phi_3 = \vec{k}_{\mathrm{eff}}.\vec{a}\,T^2$, where $\phi_i$ is the phase difference between the two counterpropagating Raman lasers at the position of the atoms at the $i$-th Raman pulse. The delay between the release of the atoms from the trap and the first Raman pulse is $t_0 = 3$ ms. The Raman beams have a waist of 5.5 mm ($1/e^2$ radius).
Thus, due to the free fall of the atoms across the laser beam, our interrogation time is limited and the intensity seen by the atoms for the three laser pulses will be different. Consequently, we adjust the timing of our experiment and the position of the laser beam in order to have the same intensity seen by the atoms for the first and the last Raman pulses. This leads to an interrogation time $2T=31.7$ ms. This configuration enables to minimize light-shift effects. As the intensity is higher for the middle pulse ($\pi$), this configuration allows also to apply the same pulse duration ($\tau=10\,\mu$s) for the three Raman laser pulses without losing too much contrast.

\begin{figure}[tbph]
\centering
\includegraphics[scale=0.35]{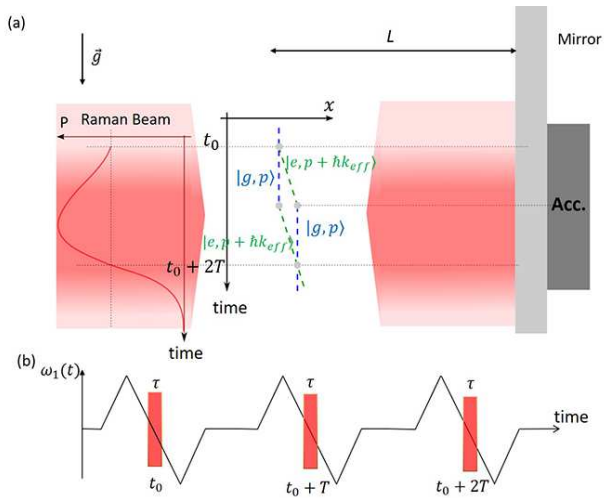}
\caption{(a) Sketch of the horizontal atomic accelerometer. The atoms fall under gravity in the retroreflected Raman beam. By measuring the transition probability $P$ as a function of free-fall time of the atoms we adjust the timing of our experiment and the position of the Raman laser beam. The pulse duration is equal for the three Raman light-pulses. The center of the classical accelerometer (Acc.) matches the position of the cold atom cloud at the $\pi$ pulse. $L = 24$ cm: atom-mirror distance. (b) Raman frequency $\omega_1(t)$ during the interferometer sequence. The same linear chirp is applied for the three Raman pulses.
}
\label{fig:figure5}
\end{figure}

\begin{figure}[tbph]
\centering
\includegraphics[scale=0.45]{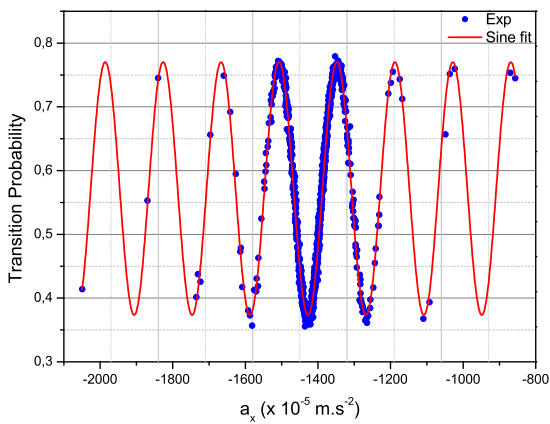}
\caption{Horizontal atom interferometer fringe pattern. The total interferometer time is $2T = 31.7$ ms and the frequency chirp applied during each Raman pulse is set to $\beta= -210$ MHz.$\mu$s$^{-1}$. The solid line is a sinusoidal least-squares fit using Eq.\ref{eq:eq2}. The estimated fringe contrast is $C \sim 40\%$.}
\label{fig:figure6}
\end{figure}

Consequently, we ensure the frequency chirp to be the same for each Raman pulse, $\beta = -210$ MHz.$\mu$s$^{-1}$ in our experiment. Figure \ref{fig:figure5} is a sketch of the interferometer setup. 
After the interferometer sequence we measure the proportion of atoms in each ouput port of the AI by fluorescence. The normalized proportion of atoms in the hyperfine state $\ket{F = 2, m_F = 0}$ after the final $\pi/2$ pulse is a sinusoidal function of the phase shift:
\begin{equation}
P=P_m-\frac{C}{2}\cos (\vec{k}_{\mathrm{eff}}.\vec{a}\,T^2)
\label{eq:eq2}
\end{equation}
where $P_m$ is the fringe offset and $C$ the fringe contrast. In a retroreflected geometry, the phase is sensitive to the acceleration of the atom compared to the mirror. Thus, in absence of vibration isolation, fluctuations of the mirror position can induce fluctuations of the interferometer phase wich wash out the fringe visibilty, even in the laboratory environment. To observe interference fringes, we perform a correlation-based technique \cite{Lautier2014} combining the simultaneous measurements of the output signal $P$ of our interferometer with the one from a classical accelerometer (QA 750, Honeywell) rigidely fixed to the Raman mirror. This allows to recover the interference fringes, although the fringes are randomly scanned by vibrations. Figure \ref{fig:figure6} displays retrieval of the fringe pattern obtained by plotting the probability transition of the AI output versus the acceleration measured by the classical accelerometer. The fringe contrast obtained from the sinusoidal least-squares fit of the data is $C = 40\%$.

\begin{figure}[tbph]
\centering
\includegraphics[scale=0.4]{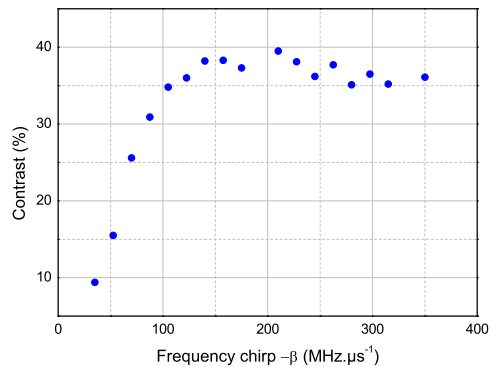}
\caption{Contrast as a function of the frequency chirp.}
\label{fig:figure7}
\end{figure}

In order to work at best sensitivity we studied the contrast of the interferometer as a function of the frequency chirp $\beta$ applied on the Raman lasers. Results are displayed on Figure \ref{fig:figure7}. The contrast is an increasing function of the frequency chirp $\beta$ until it reaches an optimum of $C = 40\%$ for $\beta=-200$ MHz.$\mu$s$^{-1}$. Increasing further the chirp value does not improve the interferometer's contrast.

\subsection{Atom accelerometer sensitivity}

\begin{figure}[tbph]
\centering
\includegraphics[scale=0.45]{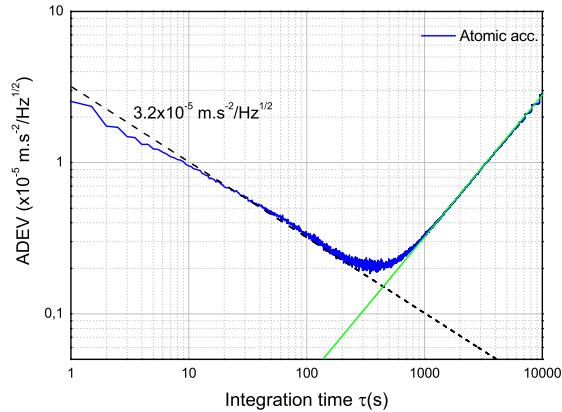}
\caption{Allan standard deviation (ADEV) of the atomic accelerometer (blue line). The dash line illustrates the $\tau^{-1/2}$ scaling. The green line illustrates the $\tau$ scaling.
}
\label{fig:figure8}
\end{figure}

To analyze the sensitivity and the stability of the horizontal atom accelerometer, we performed a hybridization of the classical accelerometer with the atom interferometer. We use the hybridization procedure described in \cite{Bidel2018}. We have operated the atomic sensor continuously during one night. Figure \ref{fig:figure8} displays the Allan standard deviation (ADEV) of the hybridized atomic accelerometer signal. The sign of the effective Raman wave vector $\vec{k}_{\mathrm{eff}}$ is reversed every measurement cycle. We achieve a short-term sensitivity of $3.2\times 10^{-5}$ m.s$^{-2}$/$\sqrt{Hz}$ which is comparable to state-of-the-art \cite{Xu2017} ($1\times 10^{-5}$ m.s$^{-2}$/$\sqrt{Hz}$) despite the use of a shorter interrogation time ($31.7$ ms versus $226$ ms). The ADEV of the horizontal acceleration measurement scales as $\tau^{-1/2}$ and reaches $0.2\times 10^{-5}$ m.s$^{-2}$ at 500 s integration time. For longer integration times, the acceleration measurement drifts as illustrated by the typical linear dependance in the averaging time $\tau$. 
The observed drift could be caused by an angular variation of the Raman mirror. The atom interferometer is measuring the projection of the gravity along the normal of the mirror. An angular drift of the mirror of 10 $\mu$rad, which seems reasonnable on our experimental setup, could explain the oberved drift.
Thus, one cannot conclude on the long term stability of the atom accelerometer unless using an auxilliary tilt sensor to monitor the angle between the Raman beam and the horizontal plane during the measurement.

\subsection{Bias arising from the frequency chirp}
\begin{figure}[tbph]
\centering
\includegraphics[scale=0.4]{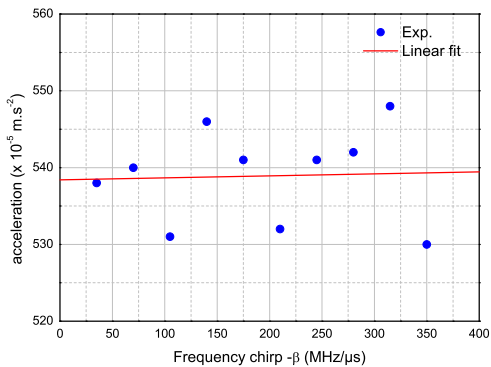}
\caption{ Acceleration as a function of the chirp $\beta$ applied on the Raman lasers. The line (in red) is a linear fit to the data points. The slope is $0.3\pm 2\times 10^{-7}$ m.s$^{-2}$/MHz.$\mu$s$^{-1}$.
}
\label{fig:figure9}
\end{figure}

To conclude our study, we investigated a possible bias induced by the frequency chirp $\beta$ on the acceleration measurement. In principle, as long as the chirp applied to the Raman lasers is the same for the three light-pulses, their should be no supplementary bias. Figure \ref{fig:figure9} displays the acceleration signal measured by the hybridized atom accelerometer as a function of the chirp $\beta$.
Each data point is obtained after an averaging time $\tau = 500$ s. The data are linear fitted and no significant slope is obtained. From the fit uncertainty on the slope, one can estimate a maximum bias of $4.6\times 10^{-5}$ m.s$^{-2}$ for a frequency chirp $\beta = -210$ MHz.$\mu$s$^{-1}$.

\section{Conclusion}
We have presented here an experimental demonstration of a method to adress a single counterpropagating Raman transition in a retroreflected configuration despite zero Doppler shift. Using this method we have achieved an horizontal acceleration sensitivity of $3.2\times 10^{-5}$ m.s$^{-2}$/$\sqrt{Hz}$ with a falling distance of $5.9$ mm, which is competitive with state-of-the-art \cite{Xu2017}. Improving the atom accelerometer sensitivity could be simply achieved by using larger laser beam radius combined with higher optical power, and a faster cycling rate. We have shown that no significant bias was introduced using this method. Further work would be required to assess long term stability of the atomic sensor. This method can easilly be extended to other AI configurations involving four or more pulses \cite{Dubetsky2006,Cadoret2016}, as for example in a double-loop geometry with pulse sequence $(\pi/2-\pi-\pi-\pi/2)$ for rotation measurements independant of acceleration \cite{Dutta2016}. Finally,this method appears suited for multiaxis inertial sensing using cold atoms without need for tilted laser beams \cite{Canuel2006,Wu2017}, as well as for compact tiltmeters or for experiments in microgravity environment using atom interferometers based on a single diffraction process \cite{Geiger2011,Barrett2016}.

.




%


\begin{acknowledgments}
We thank F. Nez, from Laboratoire Kastler Brossel (LKB), for helpful discussions.
M. C acknowledges funding from ONERA through research project CHAMO (Central Hybride AtoMique de l'Onera).

\end{acknowledgments}

\appendix





\bibliography{bibliography_of_article2}

\begin{thebibliography}{44}%
\makeatletter
\providecommand \@ifxundefined [1]{%
 \@ifx{#1\undefined}
}%
\providecommand \@ifnum [1]{%
 \ifnum #1\expandafter \@firstoftwo
 \else \expandafter \@secondoftwo
 \fi
}%
\providecommand \@ifx [1]{%
 \ifx #1\expandafter \@firstoftwo
 \else \expandafter \@secondoftwo
 \fi
}%
\providecommand \natexlab [1]{#1}%
\providecommand \enquote  [1]{``#1''}%
\providecommand \bibnamefont  [1]{#1}%
\providecommand \bibfnamefont [1]{#1}%
\providecommand \citenamefont [1]{#1}%
\providecommand \href@noop [0]{\@secondoftwo}%
\providecommand \href [0]{\begingroup \@sanitize@url \@href}%
\providecommand \@href[1]{\@@startlink{#1}\@@href}%
\providecommand \@@href[1]{\endgroup#1\@@endlink}%
\providecommand \@sanitize@url [0]{\catcode `\\12\catcode `\$12\catcode
  `\&12\catcode `\#12\catcode `\^12\catcode `\_12\catcode `\%12\relax}%
\providecommand \@@startlink[1]{}%
\providecommand \@@endlink[0]{}%
\providecommand \url  [0]{\begingroup\@sanitize@url \@url }%
\providecommand \@url [1]{\endgroup\@href {#1}{\urlprefix }}%
\providecommand \urlprefix  [0]{URL }%
\providecommand \Eprint [0]{\href }%
\providecommand \doibase [0]{http://dx.doi.org/}%
\providecommand \selectlanguage [0]{\@gobble}%
\providecommand \bibinfo  [0]{\@secondoftwo}%
\providecommand \bibfield  [0]{\@secondoftwo}%
\providecommand \translation [1]{[#1]}%
\providecommand \BibitemOpen [0]{}%
\providecommand \bibitemStop [0]{}%
\providecommand \bibitemNoStop [0]{.\EOS\space}%
\providecommand \EOS [0]{\spacefactor3000\relax}%
\providecommand \BibitemShut  [1]{\csname bibitem#1\endcsname}%
\let\auto@bib@innerbib\@empty
\bibitem [{\citenamefont {Peters}\ \emph {et~al.}(2001)\citenamefont {Peters},
  \citenamefont {Chung},\ and\ \citenamefont {Chu}}]{Peters2001}%
  \BibitemOpen
  \bibfield  {author} {\bibinfo {author} {\bibfnamefont {A.}~\bibnamefont
  {Peters}}, \bibinfo {author} {\bibfnamefont {K.~Y.}\ \bibnamefont {Chung}}, \
  and\ \bibinfo {author} {\bibfnamefont {S.}~\bibnamefont {Chu}},\ }\href@noop
  {} {\bibfield  {journal} {\bibinfo  {journal} {Metrologia}\ }\textbf
  {\bibinfo {volume} {38}},\ \bibinfo {pages} {25} (\bibinfo {year}
  {2001})}\BibitemShut {NoStop}%
\bibitem [{\citenamefont {Gillot}\ \emph {et~al.}(2014)\citenamefont {Gillot},
  \citenamefont {Francis}, \citenamefont {Landragin}, \citenamefont {Pereira
  Dos~Santos},\ and\ \citenamefont {Merlet}}]{Gillot2014}%
  \BibitemOpen
  \bibfield  {author} {\bibinfo {author} {\bibfnamefont {P.}~\bibnamefont
  {Gillot}}, \bibinfo {author} {\bibfnamefont {O.}~\bibnamefont {Francis}},
  \bibinfo {author} {\bibfnamefont {A.}~\bibnamefont {Landragin}}, \bibinfo
  {author} {\bibfnamefont {F.}~\bibnamefont {Pereira Dos~Santos}}, \ and\
  \bibinfo {author} {\bibfnamefont {S.}~\bibnamefont {Merlet}},\ }\href@noop {}
  {\bibfield  {journal} {\bibinfo  {journal} {Metrologia}\ }\textbf {\bibinfo
  {volume} {51}},\ \bibinfo {pages} {15} (\bibinfo {year} {2014})}\BibitemShut
  {NoStop}%
\bibitem [{\citenamefont {Hu}\ \emph {et~al.}(2013)\citenamefont {Hu},
  \citenamefont {Sun}, \citenamefont {Duan}, \citenamefont {Zhou},
  \citenamefont {Chen}, \citenamefont {Zhan}, \citenamefont {Zhang},\ and\
  \citenamefont {Luo}}]{Hu2013}%
  \BibitemOpen
  \bibfield  {author} {\bibinfo {author} {\bibfnamefont {Z.~K.}\ \bibnamefont
  {Hu}}, \bibinfo {author} {\bibfnamefont {B.~L.}\ \bibnamefont {Sun}},
  \bibinfo {author} {\bibfnamefont {X.~C.}\ \bibnamefont {Duan}}, \bibinfo
  {author} {\bibfnamefont {M.~K.}\ \bibnamefont {Zhou}}, \bibinfo {author}
  {\bibfnamefont {L.~L.}\ \bibnamefont {Chen}}, \bibinfo {author}
  {\bibfnamefont {S.}~\bibnamefont {Zhan}}, \bibinfo {author} {\bibfnamefont
  {Q.~Z.}\ \bibnamefont {Zhang}}, \ and\ \bibinfo {author} {\bibfnamefont
  {J.}~\bibnamefont {Luo}},\ }\href@noop {} {\bibfield  {journal} {\bibinfo
  {journal} {Physical Review A}\ }\textbf {\bibinfo {volume} {88}},\ \bibinfo
  {pages} {043610} (\bibinfo {year} {2013})}\BibitemShut {NoStop}%
\bibitem [{\citenamefont {Bidel}\ \emph {et~al.}(2018)\citenamefont {Bidel},
  \citenamefont {Zahzam}, \citenamefont {Blanchard}, \citenamefont {Bonnin},
  \citenamefont {Cadoret}, \citenamefont {Bresson}, \citenamefont {Rouxel},\
  and\ \citenamefont {Lequentrec-Lalancette}}]{Bidel2018}%
  \BibitemOpen
  \bibfield  {author} {\bibinfo {author} {\bibfnamefont {Y.}~\bibnamefont
  {Bidel}}, \bibinfo {author} {\bibfnamefont {N.}~\bibnamefont {Zahzam}},
  \bibinfo {author} {\bibfnamefont {C.}~\bibnamefont {Blanchard}}, \bibinfo
  {author} {\bibfnamefont {A.}~\bibnamefont {Bonnin}}, \bibinfo {author}
  {\bibfnamefont {M.}~\bibnamefont {Cadoret}}, \bibinfo {author} {\bibfnamefont
  {A.}~\bibnamefont {Bresson}}, \bibinfo {author} {\bibfnamefont
  {D.}~\bibnamefont {Rouxel}}, \ and\ \bibinfo {author} {\bibfnamefont {M.-F.}\
  \bibnamefont {Lequentrec-Lalancette}},\ }\href@noop {} {\bibfield  {journal}
  {\bibinfo  {journal} {Nat. Commun.}\ }\textbf {\bibinfo {volume} {9}},\
  \bibinfo {pages} {627} (\bibinfo {year} {2018})}\BibitemShut {NoStop}%
\bibitem [{\citenamefont {Hauth}\ \emph {et~al.}(2013)\citenamefont {Hauth},
  \citenamefont {Freier}, \citenamefont {Schkolnik}, \citenamefont {Senger},
  \citenamefont {Schmidt},\ and\ \citenamefont {Peters}}]{Hauth2013}%
  \BibitemOpen
  \bibfield  {author} {\bibinfo {author} {\bibfnamefont {M.}~\bibnamefont
  {Hauth}}, \bibinfo {author} {\bibfnamefont {C.}~\bibnamefont {Freier}},
  \bibinfo {author} {\bibfnamefont {V.}~\bibnamefont {Schkolnik}}, \bibinfo
  {author} {\bibfnamefont {A.}~\bibnamefont {Senger}}, \bibinfo {author}
  {\bibfnamefont {M.}~\bibnamefont {Schmidt}}, \ and\ \bibinfo {author}
  {\bibfnamefont {A.}~\bibnamefont {Peters}},\ }\href@noop {} {\bibfield
  {journal} {\bibinfo  {journal} {Applied Physics B}\ }\textbf {\bibinfo
  {volume} {113}},\ \bibinfo {pages} {49} (\bibinfo {year} {2013})}\BibitemShut
  {NoStop}%
\bibitem [{\citenamefont {McGuirk}\ \emph {et~al.}(2002)\citenamefont
  {McGuirk}, \citenamefont {Foster}, \citenamefont {Fixler}, \citenamefont
  {Snadden},\ and\ \citenamefont {Kasevich}}]{McGuirk2002}%
  \BibitemOpen
  \bibfield  {author} {\bibinfo {author} {\bibfnamefont {J.~M.}\ \bibnamefont
  {McGuirk}}, \bibinfo {author} {\bibfnamefont {G.~T.}\ \bibnamefont {Foster}},
  \bibinfo {author} {\bibfnamefont {J.~B.}\ \bibnamefont {Fixler}}, \bibinfo
  {author} {\bibfnamefont {M.~J.}\ \bibnamefont {Snadden}}, \ and\ \bibinfo
  {author} {\bibfnamefont {M.~A.}\ \bibnamefont {Kasevich}},\ }\href@noop {}
  {\bibfield  {journal} {\bibinfo  {journal} {Phys. Rev. A}\ }\textbf {\bibinfo
  {volume} {65}},\ \bibinfo {pages} {033608} (\bibinfo {year}
  {2002})}\BibitemShut {NoStop}%
\bibitem [{\citenamefont {Sorrentino}\ \emph {et~al.}(2014)\citenamefont
  {Sorrentino}, \citenamefont {Bodart}, \citenamefont {Cacciapuoti},
  \citenamefont {Lien}, \citenamefont {Prevedelli}, \citenamefont {Rosi},
  \citenamefont {Salvi},\ and\ \citenamefont {Tino}}]{Sorrentino2014}%
  \BibitemOpen
  \bibfield  {author} {\bibinfo {author} {\bibfnamefont {F.}~\bibnamefont
  {Sorrentino}}, \bibinfo {author} {\bibfnamefont {Q.}~\bibnamefont {Bodart}},
  \bibinfo {author} {\bibfnamefont {L.}~\bibnamefont {Cacciapuoti}}, \bibinfo
  {author} {\bibfnamefont {Y.-H.}\ \bibnamefont {Lien}}, \bibinfo {author}
  {\bibfnamefont {M.}~\bibnamefont {Prevedelli}}, \bibinfo {author}
  {\bibfnamefont {G.}~\bibnamefont {Rosi}}, \bibinfo {author} {\bibfnamefont
  {L.}~\bibnamefont {Salvi}}, \ and\ \bibinfo {author} {\bibfnamefont {G.~M.}\
  \bibnamefont {Tino}},\ }\href@noop {} {\bibfield  {journal} {\bibinfo
  {journal} {Phys. Rev. A}\ }\textbf {\bibinfo {volume} {89}},\ \bibinfo
  {pages} {023607} (\bibinfo {year} {2014})}\BibitemShut {NoStop}%
\bibitem [{\citenamefont {Duan}\ \emph {et~al.}(2014)\citenamefont {Duan},
  \citenamefont {Zhou}, \citenamefont {Mao}, \citenamefont {Yao}, \citenamefont
  {Deng}, \citenamefont {Luo},\ and\ \citenamefont {Hu}}]{Duan2014}%
  \BibitemOpen
  \bibfield  {author} {\bibinfo {author} {\bibfnamefont {X.-C.}\ \bibnamefont
  {Duan}}, \bibinfo {author} {\bibfnamefont {M.-K.}\ \bibnamefont {Zhou}},
  \bibinfo {author} {\bibfnamefont {D.-K.}\ \bibnamefont {Mao}}, \bibinfo
  {author} {\bibfnamefont {H.-B.}\ \bibnamefont {Yao}}, \bibinfo {author}
  {\bibfnamefont {X.-B.}\ \bibnamefont {Deng}}, \bibinfo {author}
  {\bibfnamefont {J.}~\bibnamefont {Luo}}, \ and\ \bibinfo {author}
  {\bibfnamefont {Z.-K.}\ \bibnamefont {Hu}},\ }\href@noop {} {\bibfield
  {journal} {\bibinfo  {journal} {Phys. Rev. A}\ }\textbf {\bibinfo {volume}
  {90}},\ \bibinfo {pages} {023617} (\bibinfo {year} {2014})}\BibitemShut
  {NoStop}%
\bibitem [{\citenamefont {Gustavson}\ \emph {et~al.}(1997)\citenamefont
  {Gustavson}, \citenamefont {Bouyer},\ and\ \citenamefont
  {Kasevich}}]{Gustavson1997}%
  \BibitemOpen
  \bibfield  {author} {\bibinfo {author} {\bibfnamefont {T.~L.}\ \bibnamefont
  {Gustavson}}, \bibinfo {author} {\bibfnamefont {P.}~\bibnamefont {Bouyer}}, \
  and\ \bibinfo {author} {\bibfnamefont {M.~A.}\ \bibnamefont {Kasevich}},\
  }\href@noop {} {\bibfield  {journal} {\bibinfo  {journal} {Phys. Rev. Lett.}\
  }\textbf {\bibinfo {volume} {78}},\ \bibinfo {pages} {2046} (\bibinfo {year}
  {1997})}\BibitemShut {NoStop}%
\bibitem [{\citenamefont {Gauguet}\ \emph {et~al.}(2009)\citenamefont
  {Gauguet}, \citenamefont {Canuel}, \citenamefont {L\'ev\`eque}, \citenamefont
  {Chaibi},\ and\ \citenamefont {Landragin}}]{Gauguet2009}%
  \BibitemOpen
  \bibfield  {author} {\bibinfo {author} {\bibfnamefont {A.}~\bibnamefont
  {Gauguet}}, \bibinfo {author} {\bibfnamefont {B.}~\bibnamefont {Canuel}},
  \bibinfo {author} {\bibfnamefont {T.}~\bibnamefont {L\'ev\`eque}}, \bibinfo
  {author} {\bibfnamefont {W.}~\bibnamefont {Chaibi}}, \ and\ \bibinfo {author}
  {\bibfnamefont {A.}~\bibnamefont {Landragin}},\ }\href@noop {} {\bibfield
  {journal} {\bibinfo  {journal} {Phys. Rev. A}\ }\textbf {\bibinfo {volume}
  {80}},\ \bibinfo {pages} {063604} (\bibinfo {year} {2009})}\BibitemShut
  {NoStop}%
\bibitem [{\citenamefont {Tackmann}\ \emph {et~al.}(2012)\citenamefont
  {Tackmann}, \citenamefont {Berg}, \citenamefont {Schubert}, \citenamefont
  {Abend}, \citenamefont {Gilowski}, \citenamefont {Ertmer},\ and\
  \citenamefont {Rasel}}]{Tackmann2012}%
  \BibitemOpen
  \bibfield  {author} {\bibinfo {author} {\bibfnamefont {G.}~\bibnamefont
  {Tackmann}}, \bibinfo {author} {\bibfnamefont {P.}~\bibnamefont {Berg}},
  \bibinfo {author} {\bibfnamefont {C.}~\bibnamefont {Schubert}}, \bibinfo
  {author} {\bibfnamefont {S.}~\bibnamefont {Abend}}, \bibinfo {author}
  {\bibfnamefont {M.}~\bibnamefont {Gilowski}}, \bibinfo {author}
  {\bibfnamefont {W.}~\bibnamefont {Ertmer}}, \ and\ \bibinfo {author}
  {\bibfnamefont {E.~M.}\ \bibnamefont {Rasel}},\ }\href@noop {} {\bibfield
  {journal} {\bibinfo  {journal} {New Journal of Physics}\ }\textbf {\bibinfo
  {volume} {14}},\ \bibinfo {pages} {015002} (\bibinfo {year}
  {2012})}\BibitemShut {NoStop}%
\bibitem [{\citenamefont {Dutta}\ \emph {et~al.}(2016)\citenamefont {Dutta},
  \citenamefont {Savoie}, \citenamefont {Fang}, \citenamefont {Venon},
  \citenamefont {Alzar}, \citenamefont {Geiger},\ and\ \citenamefont
  {Landragin}}]{Dutta2016}%
  \BibitemOpen
  \bibfield  {author} {\bibinfo {author} {\bibfnamefont {I.}~\bibnamefont
  {Dutta}}, \bibinfo {author} {\bibfnamefont {D.}~\bibnamefont {Savoie}},
  \bibinfo {author} {\bibfnamefont {B.}~\bibnamefont {Fang}}, \bibinfo {author}
  {\bibfnamefont {B.}~\bibnamefont {Venon}}, \bibinfo {author} {\bibfnamefont
  {C.~L.~G.}\ \bibnamefont {Alzar}}, \bibinfo {author} {\bibfnamefont
  {R.}~\bibnamefont {Geiger}}, \ and\ \bibinfo {author} {\bibfnamefont
  {A.}~\bibnamefont {Landragin}},\ }\href@noop {} {\bibfield  {journal}
  {\bibinfo  {journal} {Phys. Rev. Lett.}\ }\textbf {\bibinfo {volume} {116}},\
  \bibinfo {pages} {183003} (\bibinfo {year} {2016})}\BibitemShut {NoStop}%
\bibitem [{\citenamefont {Jekeli}(2005)}]{Jekeli2005}%
  \BibitemOpen
  \bibfield  {author} {\bibinfo {author} {\bibfnamefont {C.}~\bibnamefont
  {Jekeli}},\ }\href@noop {} {\bibfield  {journal} {\bibinfo  {journal}
  {Navigation}\ }\textbf {\bibinfo {volume} {2}},\ \bibinfo {pages} {1}
  (\bibinfo {year} {2005})}\BibitemShut {NoStop}%
\bibitem [{\citenamefont {Parker}\ \emph {et~al.}(2018)\citenamefont {Parker},
  \citenamefont {Yu}, \citenamefont {Zhong}, \citenamefont {Estey},\ and\
  \citenamefont {M\"uller}}]{Muller2018}%
  \BibitemOpen
  \bibfield  {author} {\bibinfo {author} {\bibfnamefont {R.~H.}\ \bibnamefont
  {Parker}}, \bibinfo {author} {\bibfnamefont {C.}~\bibnamefont {Yu}}, \bibinfo
  {author} {\bibfnamefont {W.}~\bibnamefont {Zhong}}, \bibinfo {author}
  {\bibfnamefont {B.}~\bibnamefont {Estey}}, \ and\ \bibinfo {author}
  {\bibfnamefont {H.}~\bibnamefont {M\"uller}},\ }\href@noop {} {\bibfield
  {journal} {\bibinfo  {journal} {Science}\ }\textbf {\bibinfo {volume}
  {360}},\ \bibinfo {pages} {191} (\bibinfo {year} {2018})}\BibitemShut
  {NoStop}%
\bibitem [{\citenamefont {Fixler}\ \emph {et~al.}(2007)\citenamefont {Fixler},
  \citenamefont {Foster}, \citenamefont {McGuirk},\ and\ \citenamefont
  {Kasevich}}]{Fixler2007}%
  \BibitemOpen
  \bibfield  {author} {\bibinfo {author} {\bibfnamefont {J.~B.}\ \bibnamefont
  {Fixler}}, \bibinfo {author} {\bibfnamefont {G.~T.}\ \bibnamefont {Foster}},
  \bibinfo {author} {\bibfnamefont {J.~M.}\ \bibnamefont {McGuirk}}, \ and\
  \bibinfo {author} {\bibfnamefont {M.~A.}\ \bibnamefont {Kasevich}},\
  }\href@noop {} {\bibfield  {journal} {\bibinfo  {journal} {Science}\ }\textbf
  {\bibinfo {volume} {315}},\ \bibinfo {pages} {74} (\bibinfo {year}
  {2007})}\BibitemShut {NoStop}%
\bibitem [{\citenamefont {Bouchendira}\ \emph {et~al.}(2011)\citenamefont
  {Bouchendira}, \citenamefont {Clad\'{e}}, \citenamefont
  {Guellati-Kh\'{e}lifa}, \citenamefont {Nez},\ and\ \citenamefont
  {Biraben}}]{Bouchendira2011}%
  \BibitemOpen
  \bibfield  {author} {\bibinfo {author} {\bibfnamefont {R.}~\bibnamefont
  {Bouchendira}}, \bibinfo {author} {\bibfnamefont {P.}~\bibnamefont
  {Clad\'{e}}}, \bibinfo {author} {\bibfnamefont {S.}~\bibnamefont
  {Guellati-Kh\'{e}lifa}}, \bibinfo {author} {\bibfnamefont {F.}~\bibnamefont
  {Nez}}, \ and\ \bibinfo {author} {\bibfnamefont {F.}~\bibnamefont
  {Biraben}},\ }\href@noop {} {\bibfield  {journal} {\bibinfo  {journal} {Phys.
  Rev. Lett.}\ }\textbf {\bibinfo {volume} {106}},\ \bibinfo {pages} {080801}
  (\bibinfo {year} {2011})}\BibitemShut {NoStop}%
\bibitem [{\citenamefont {Rosi}\ \emph {et~al.}(2014)\citenamefont {Rosi},
  \citenamefont {Sorrentino}, \citenamefont {Cacciapuoti}, \citenamefont
  {Prevedelli},\ and\ \citenamefont {Tino}}]{Rosi2014}%
  \BibitemOpen
  \bibfield  {author} {\bibinfo {author} {\bibfnamefont {G.}~\bibnamefont
  {Rosi}}, \bibinfo {author} {\bibfnamefont {F.}~\bibnamefont {Sorrentino}},
  \bibinfo {author} {\bibfnamefont {L.}~\bibnamefont {Cacciapuoti}}, \bibinfo
  {author} {\bibfnamefont {M.}~\bibnamefont {Prevedelli}}, \ and\ \bibinfo
  {author} {\bibfnamefont {G.~M.}\ \bibnamefont {Tino}},\ }\href@noop {}
  {\bibfield  {journal} {\bibinfo  {journal} {Nature}\ }\textbf {\bibinfo
  {volume} {510}},\ \bibinfo {pages} {518} (\bibinfo {year}
  {2014})}\BibitemShut {NoStop}%
\bibitem [{\citenamefont {Fray}\ \emph {et~al.}(2004)\citenamefont {Fray},
  \citenamefont {Diez}, \citenamefont {H\"ansch},\ and\ \citenamefont
  {Weitz}}]{Fray2004}%
  \BibitemOpen
  \bibfield  {author} {\bibinfo {author} {\bibfnamefont {S.}~\bibnamefont
  {Fray}}, \bibinfo {author} {\bibfnamefont {C.~A.}\ \bibnamefont {Diez}},
  \bibinfo {author} {\bibfnamefont {T.~W.}\ \bibnamefont {H\"ansch}}, \ and\
  \bibinfo {author} {\bibfnamefont {M.}~\bibnamefont {Weitz}},\ }\href@noop {}
  {\bibfield  {journal} {\bibinfo  {journal} {Phys. Rev. Lett.}\ }\textbf
  {\bibinfo {volume} {93}},\ \bibinfo {pages} {240404} (\bibinfo {year}
  {2004})}\BibitemShut {NoStop}%
\bibitem [{\citenamefont {Bonnin}\ \emph {et~al.}(2013)\citenamefont {Bonnin},
  \citenamefont {Zahzam}, \citenamefont {Bidel},\ and\ \citenamefont
  {Bresson}}]{Bonnin2013}%
  \BibitemOpen
  \bibfield  {author} {\bibinfo {author} {\bibfnamefont {A.}~\bibnamefont
  {Bonnin}}, \bibinfo {author} {\bibfnamefont {N.}~\bibnamefont {Zahzam}},
  \bibinfo {author} {\bibfnamefont {Y.}~\bibnamefont {Bidel}}, \ and\ \bibinfo
  {author} {\bibfnamefont {A.}~\bibnamefont {Bresson}},\ }\href@noop {}
  {\bibfield  {journal} {\bibinfo  {journal} {Phys. Rev. A}\ }\textbf {\bibinfo
  {volume} {88}},\ \bibinfo {pages} {043615} (\bibinfo {year}
  {2013})}\BibitemShut {NoStop}%
\bibitem [{\citenamefont {Schlippert}\ \emph {et~al.}(2014)\citenamefont
  {Schlippert}, \citenamefont {Hartwig}, \citenamefont {Albers}, \citenamefont
  {Richardson}, \citenamefont {Schubert}, \citenamefont {Roura}, \citenamefont
  {Schleich}, \citenamefont {Ertmer},\ and\ \citenamefont
  {Rasel}}]{Schlippert2014}%
  \BibitemOpen
  \bibfield  {author} {\bibinfo {author} {\bibfnamefont {D.}~\bibnamefont
  {Schlippert}}, \bibinfo {author} {\bibfnamefont {J.}~\bibnamefont {Hartwig}},
  \bibinfo {author} {\bibfnamefont {H.}~\bibnamefont {Albers}}, \bibinfo
  {author} {\bibfnamefont {L.~L.}\ \bibnamefont {Richardson}}, \bibinfo
  {author} {\bibfnamefont {C.}~\bibnamefont {Schubert}}, \bibinfo {author}
  {\bibfnamefont {A.}~\bibnamefont {Roura}}, \bibinfo {author} {\bibfnamefont
  {W.~P.}\ \bibnamefont {Schleich}}, \bibinfo {author} {\bibfnamefont
  {W.}~\bibnamefont {Ertmer}}, \ and\ \bibinfo {author} {\bibfnamefont {E.~M.}\
  \bibnamefont {Rasel}},\ }\href@noop {} {\bibfield  {journal} {\bibinfo
  {journal} {Phys. Rev. Lett.}\ }\textbf {\bibinfo {volume} {112}},\ \bibinfo
  {pages} {203002} (\bibinfo {year} {2014})}\BibitemShut {NoStop}%
\bibitem [{\citenamefont {Tarallo}\ \emph {et~al.}(2014)\citenamefont
  {Tarallo}, \citenamefont {Mazzoni}, \citenamefont {Poli}, \citenamefont
  {Sutyrin}, \citenamefont {Zhang},\ and\ \citenamefont {Tino}}]{Tarallo2014}%
  \BibitemOpen
  \bibfield  {author} {\bibinfo {author} {\bibfnamefont {M.~G.}\ \bibnamefont
  {Tarallo}}, \bibinfo {author} {\bibfnamefont {T.}~\bibnamefont {Mazzoni}},
  \bibinfo {author} {\bibfnamefont {N.}~\bibnamefont {Poli}}, \bibinfo {author}
  {\bibfnamefont {D.~V.}\ \bibnamefont {Sutyrin}}, \bibinfo {author}
  {\bibfnamefont {X.}~\bibnamefont {Zhang}}, \ and\ \bibinfo {author}
  {\bibfnamefont {G.~M.}\ \bibnamefont {Tino}},\ }\href@noop {} {\bibfield
  {journal} {\bibinfo  {journal} {Phys. Rev. Lett.}\ }\textbf {\bibinfo
  {volume} {113}},\ \bibinfo {pages} {023005} (\bibinfo {year}
  {2014})}\BibitemShut {NoStop}%
\bibitem [{\citenamefont {Zhou}\ \emph {et~al.}(2015)\citenamefont {Zhou},
  \citenamefont {Long}, \citenamefont {Tang}, \citenamefont {Chen},
  \citenamefont {Gao}, \citenamefont {Peng}, \citenamefont {Duan},\ and\
  \citenamefont {Zhong}}]{Zhou2015}%
  \BibitemOpen
  \bibfield  {author} {\bibinfo {author} {\bibfnamefont {L.}~\bibnamefont
  {Zhou}}, \bibinfo {author} {\bibfnamefont {S.}~\bibnamefont {Long}}, \bibinfo
  {author} {\bibfnamefont {B.}~\bibnamefont {Tang}}, \bibinfo {author}
  {\bibfnamefont {X.}~\bibnamefont {Chen}}, \bibinfo {author} {\bibfnamefont
  {F.}~\bibnamefont {Gao}}, \bibinfo {author} {\bibfnamefont {W.}~\bibnamefont
  {Peng}}, \bibinfo {author} {\bibfnamefont {W.}~\bibnamefont {Duan}}, \ and\
  \bibinfo {author} {\bibfnamefont {J.}~\bibnamefont {Zhong}},\ }\href@noop {}
  {\bibfield  {journal} {\bibinfo  {journal} {Phys. Rev. Lett.}\ }\textbf
  {\bibinfo {volume} {115}},\ \bibinfo {pages} {013004} (\bibinfo {year}
  {2015})}\BibitemShut {NoStop}%
\bibitem [{\citenamefont {Hamilton}\ \emph {et~al.}(2015)\citenamefont
  {Hamilton}, \citenamefont {Jaffe}, \citenamefont {Haslinger}, \citenamefont
  {Simmons}, \citenamefont {M\"uller},\ and\ \citenamefont
  {Khoury}}]{Hamilton2015}%
  \BibitemOpen
  \bibfield  {author} {\bibinfo {author} {\bibfnamefont {P.}~\bibnamefont
  {Hamilton}}, \bibinfo {author} {\bibfnamefont {M.}~\bibnamefont {Jaffe}},
  \bibinfo {author} {\bibfnamefont {P.}~\bibnamefont {Haslinger}}, \bibinfo
  {author} {\bibfnamefont {Q.}~\bibnamefont {Simmons}}, \bibinfo {author}
  {\bibfnamefont {H.}~\bibnamefont {M\"uller}}, \ and\ \bibinfo {author}
  {\bibfnamefont {J.}~\bibnamefont {Khoury}},\ }\href@noop {} {\bibfield
  {journal} {\bibinfo  {journal} {Science}\ }\textbf {\bibinfo {volume}
  {349}},\ \bibinfo {pages} {849} (\bibinfo {year} {2015})}\BibitemShut
  {NoStop}%
\bibitem [{\citenamefont {Dimopoulos}\ \emph {et~al.}(2008)\citenamefont
  {Dimopoulos}, \citenamefont {Graham}, \citenamefont {Hogan}, \citenamefont
  {Kasevich},\ and\ \citenamefont {Rajendran}}]{Dimopoulos2008}%
  \BibitemOpen
  \bibfield  {author} {\bibinfo {author} {\bibfnamefont {S.}~\bibnamefont
  {Dimopoulos}}, \bibinfo {author} {\bibfnamefont {P.~W.}\ \bibnamefont
  {Graham}}, \bibinfo {author} {\bibfnamefont {J.~M.}\ \bibnamefont {Hogan}},
  \bibinfo {author} {\bibfnamefont {M.~A.}\ \bibnamefont {Kasevich}}, \ and\
  \bibinfo {author} {\bibfnamefont {S.}~\bibnamefont {Rajendran}},\ }\href@noop
  {} {\bibfield  {journal} {\bibinfo  {journal} {Phys. Rev. D}\ }\textbf
  {\bibinfo {volume} {78}},\ \bibinfo {pages} {122002} (\bibinfo {year}
  {2008})}\BibitemShut {NoStop}%
\bibitem [{\citenamefont {Chaibi}\ \emph {et~al.}(2016)\citenamefont {Chaibi},
  \citenamefont {Geiger}, \citenamefont {Canuel}, \citenamefont {Bertoldi},
  \citenamefont {Landragin},\ and\ \citenamefont {Bouyer}}]{Chaibi2016}%
  \BibitemOpen
  \bibfield  {author} {\bibinfo {author} {\bibfnamefont {W.}~\bibnamefont
  {Chaibi}}, \bibinfo {author} {\bibfnamefont {R.}~\bibnamefont {Geiger}},
  \bibinfo {author} {\bibfnamefont {B.}~\bibnamefont {Canuel}}, \bibinfo
  {author} {\bibfnamefont {A.}~\bibnamefont {Bertoldi}}, \bibinfo {author}
  {\bibfnamefont {A.}~\bibnamefont {Landragin}}, \ and\ \bibinfo {author}
  {\bibfnamefont {P.}~\bibnamefont {Bouyer}},\ }\href@noop {} {\bibfield
  {journal} {\bibinfo  {journal} {Phys. Rev. D}\ }\textbf {\bibinfo {volume}
  {93}},\ \bibinfo {pages} {021101} (\bibinfo {year} {2016})}\BibitemShut
  {NoStop}%
\bibitem [{\citenamefont {Hamilton}\ \emph {et~al.}(2014)\citenamefont
  {Hamilton}, \citenamefont {Zhmoginov}, \citenamefont {Robicheaux},
  \citenamefont {Fajans}, \citenamefont {Wurtele},\ and\ \citenamefont
  {üller}}]{Hamilton2014}%
  \BibitemOpen
  \bibfield  {author} {\bibinfo {author} {\bibfnamefont {P.}~\bibnamefont
  {Hamilton}}, \bibinfo {author} {\bibfnamefont {A.}~\bibnamefont {Zhmoginov}},
  \bibinfo {author} {\bibfnamefont {F.}~\bibnamefont {Robicheaux}}, \bibinfo
  {author} {\bibfnamefont {J.}~\bibnamefont {Fajans}}, \bibinfo {author}
  {\bibfnamefont {J.~S.}\ \bibnamefont {Wurtele}}, \ and\ \bibinfo {author}
  {\bibfnamefont {H.~M.}\ \bibnamefont {üller}},\ }\href@noop {} {\bibfield
  {journal} {\bibinfo  {journal} {Phys. Rev. Lett}\ }\textbf {\bibinfo {volume}
  {112}},\ \bibinfo {pages} {121102} (\bibinfo {year} {2014})}\BibitemShut
  {NoStop}%
\bibitem [{\citenamefont {Kasevich}\ and\ \citenamefont
  {Chu}(1991)}]{Kasevich1991}%
  \BibitemOpen
  \bibfield  {author} {\bibinfo {author} {\bibfnamefont {M.}~\bibnamefont
  {Kasevich}}\ and\ \bibinfo {author} {\bibfnamefont {S.}~\bibnamefont {Chu}},\
  }\href@noop {} {\bibfield  {journal} {\bibinfo  {journal} {Phys. Rev. Lett}\
  }\textbf {\bibinfo {volume} {67}},\ \bibinfo {pages} {181} (\bibinfo {year}
  {1991})}\BibitemShut {NoStop}%
\bibitem [{\citenamefont {Schkolnik}\ \emph {et~al.}(2015)\citenamefont
  {Schkolnik}, \citenamefont {Leykauf}, \citenamefont {Hauth}, \citenamefont
  {Freier},\ and\ \citenamefont {Peters}}]{Peters2015}%
  \BibitemOpen
  \bibfield  {author} {\bibinfo {author} {\bibfnamefont {V.}~\bibnamefont
  {Schkolnik}}, \bibinfo {author} {\bibfnamefont {B.}~\bibnamefont {Leykauf}},
  \bibinfo {author} {\bibfnamefont {M.}~\bibnamefont {Hauth}}, \bibinfo
  {author} {\bibfnamefont {C.}~\bibnamefont {Freier}}, \ and\ \bibinfo {author}
  {\bibfnamefont {A.}~\bibnamefont {Peters}},\ }\href@noop {} {\bibfield
  {journal} {\bibinfo  {journal} {App. Phys. B}\ }\textbf {\bibinfo {volume}
  {120}},\ \bibinfo {pages} {311} (\bibinfo {year} {2015})}\BibitemShut
  {NoStop}%
\bibitem [{\citenamefont {Durfee}\ \emph {et~al.}(2006)\citenamefont {Durfee},
  \citenamefont {Shaham},\ and\ \citenamefont {Kasevich}}]{Durfee2006}%
  \BibitemOpen
  \bibfield  {author} {\bibinfo {author} {\bibfnamefont {D.~S.}\ \bibnamefont
  {Durfee}}, \bibinfo {author} {\bibfnamefont {Y.~K.}\ \bibnamefont {Shaham}},
  \ and\ \bibinfo {author} {\bibfnamefont {M.~A.}\ \bibnamefont {Kasevich}},\
  }\href@noop {} {\bibfield  {journal} {\bibinfo  {journal} {Phys. Rev. Lett}\
  }\textbf {\bibinfo {volume} {97}},\ \bibinfo {pages} {240801} (\bibinfo
  {year} {2006})}\BibitemShut {NoStop}%
\bibitem [{\citenamefont {L\'ev\`eque}\ \emph {et~al.}(2009)\citenamefont
  {L\'ev\`eque}, \citenamefont {Gauguet}, \citenamefont {Michaud},
  \citenamefont {Pereira Dos~Santos},\ and\ \citenamefont
  {Landragin}}]{Leveque2009}%
  \BibitemOpen
  \bibfield  {author} {\bibinfo {author} {\bibfnamefont {T.}~\bibnamefont
  {L\'ev\`eque}}, \bibinfo {author} {\bibfnamefont {A.}~\bibnamefont
  {Gauguet}}, \bibinfo {author} {\bibfnamefont {F.}~\bibnamefont {Michaud}},
  \bibinfo {author} {\bibfnamefont {F.}~\bibnamefont {Pereira Dos~Santos}}, \
  and\ \bibinfo {author} {\bibfnamefont {A.}~\bibnamefont {Landragin}},\
  }\href@noop {} {\bibfield  {journal} {\bibinfo  {journal} {Phys. Rev. Lett.}\
  }\textbf {\bibinfo {volume} {103}},\ \bibinfo {pages} {080405} (\bibinfo
  {year} {2009})}\BibitemShut {NoStop}%
\bibitem [{\citenamefont {Malossi}\ \emph {et~al.}(2010)\citenamefont
  {Malossi}, \citenamefont {Bodart}, \citenamefont {Merlet}, \citenamefont
  {L\'ev\`eque}, \citenamefont {Landragin},\ and\ \citenamefont {Pereira
  Dos~Santos}}]{Malossi2010}%
  \BibitemOpen
  \bibfield  {author} {\bibinfo {author} {\bibfnamefont {N.}~\bibnamefont
  {Malossi}}, \bibinfo {author} {\bibfnamefont {Q.}~\bibnamefont {Bodart}},
  \bibinfo {author} {\bibfnamefont {S.}~\bibnamefont {Merlet}}, \bibinfo
  {author} {\bibfnamefont {T.}~\bibnamefont {L\'ev\`eque}}, \bibinfo {author}
  {\bibfnamefont {A.}~\bibnamefont {Landragin}}, \ and\ \bibinfo {author}
  {\bibfnamefont {F.}~\bibnamefont {Pereira Dos~Santos}},\ }\href@noop {}
  {\bibfield  {journal} {\bibinfo  {journal} {Phys. Rev. A}\ }\textbf {\bibinfo
  {volume} {81}},\ \bibinfo {pages} {013617} (\bibinfo {year}
  {2010})}\BibitemShut {NoStop}%
\bibitem [{\citenamefont {Carraz}\ \emph {et~al.}(2014)\citenamefont {Carraz},
  \citenamefont {Siemes}, \citenamefont {Massotti}, \citenamefont {Haagmans},\
  and\ \citenamefont {Silvestrin}}]{Carraz2014}%
  \BibitemOpen
  \bibfield  {author} {\bibinfo {author} {\bibfnamefont {O.}~\bibnamefont
  {Carraz}}, \bibinfo {author} {\bibfnamefont {C.}~\bibnamefont {Siemes}},
  \bibinfo {author} {\bibfnamefont {L.}~\bibnamefont {Massotti}}, \bibinfo
  {author} {\bibfnamefont {R.}~\bibnamefont {Haagmans}}, \ and\ \bibinfo
  {author} {\bibfnamefont {P.}~\bibnamefont {Silvestrin}},\ }\href@noop {}
  {\bibfield  {journal} {\bibinfo  {journal} {Microgravity Science and
  Technology}\ }\textbf {\bibinfo {volume} {26}},\ \bibinfo {pages} {139}
  (\bibinfo {year} {2014})}\BibitemShut {NoStop}%
\bibitem [{\citenamefont {Perrin}\ \emph {et~al.}(2019)\citenamefont {Perrin},
  \citenamefont {Bidel}, \citenamefont {Zahzam}, \citenamefont {Blanchard},
  \citenamefont {Bresson},\ and\ \citenamefont {Cadoret}}]{Perrin2019}%
  \BibitemOpen
  \bibfield  {author} {\bibinfo {author} {\bibfnamefont {I.}~\bibnamefont
  {Perrin}}, \bibinfo {author} {\bibfnamefont {Y.}~\bibnamefont {Bidel}},
  \bibinfo {author} {\bibfnamefont {N.}~\bibnamefont {Zahzam}}, \bibinfo
  {author} {\bibfnamefont {C.}~\bibnamefont {Blanchard}}, \bibinfo {author}
  {\bibfnamefont {A.}~\bibnamefont {Bresson}}, \ and\ \bibinfo {author}
  {\bibfnamefont {M.}~\bibnamefont {Cadoret}},\ }\href@noop {} {\bibfield
  {journal} {\bibinfo  {journal} {Phys. Rev. A}\ }\textbf {\bibinfo {volume}
  {99}},\ \bibinfo {pages} {013601} (\bibinfo {year} {2019})}\BibitemShut
  {NoStop}%
\bibitem [{\citenamefont {Th\'{e}ron}\ \emph {et~al.}(2017)\citenamefont
  {Th\'{e}ron}, \citenamefont {Bidel}, \citenamefont {Dieu}, \citenamefont
  {Zahzam}, \citenamefont {Cadoret},\ and\ \citenamefont
  {Bresson}}]{Theron2017}%
  \BibitemOpen
  \bibfield  {author} {\bibinfo {author} {\bibfnamefont {F.}~\bibnamefont
  {Th\'{e}ron}}, \bibinfo {author} {\bibfnamefont {Y.}~\bibnamefont {Bidel}},
  \bibinfo {author} {\bibfnamefont {E.}~\bibnamefont {Dieu}}, \bibinfo {author}
  {\bibfnamefont {N.}~\bibnamefont {Zahzam}}, \bibinfo {author} {\bibfnamefont
  {M.}~\bibnamefont {Cadoret}}, \ and\ \bibinfo {author} {\bibfnamefont
  {A.}~\bibnamefont {Bresson}},\ }\href@noop {} {\bibfield  {journal} {\bibinfo
   {journal} {Optics Communications}\ }\textbf {\bibinfo {volume} {393}},\
  \bibinfo {pages} {152} (\bibinfo {year} {2017})}\BibitemShut {NoStop}%
\bibitem [{\citenamefont {Th\'{e}ron}\ \emph {et~al.}(2015)\citenamefont
  {Th\'{e}ron}, \citenamefont {Carraz}, \citenamefont {Renon}, \citenamefont
  {Zahzam}, \citenamefont {Bidel}, \citenamefont {Cadoret},\ and\ \citenamefont
  {Bresson}}]{Theron2015}%
  \BibitemOpen
  \bibfield  {author} {\bibinfo {author} {\bibfnamefont {F.}~\bibnamefont
  {Th\'{e}ron}}, \bibinfo {author} {\bibfnamefont {O.}~\bibnamefont {Carraz}},
  \bibinfo {author} {\bibfnamefont {G.}~\bibnamefont {Renon}}, \bibinfo
  {author} {\bibfnamefont {N.}~\bibnamefont {Zahzam}}, \bibinfo {author}
  {\bibfnamefont {Y.}~\bibnamefont {Bidel}}, \bibinfo {author} {\bibfnamefont
  {M.}~\bibnamefont {Cadoret}}, \ and\ \bibinfo {author} {\bibfnamefont
  {A.}~\bibnamefont {Bresson}},\ }\href@noop {} {\bibfield  {journal} {\bibinfo
   {journal} {Appl. Phys. B}\ }\textbf {\bibinfo {volume} {118}},\ \bibinfo
  {pages} {1} (\bibinfo {year} {2015})}\BibitemShut {NoStop}%
\bibitem [{\citenamefont {Carraz}\ \emph {et~al.}(2012)\citenamefont {Carraz},
  \citenamefont {Charri\`{e}re}, \citenamefont {Cadoret}, \citenamefont
  {Zahzam}, \citenamefont {Bidel},\ and\ \citenamefont {Bresson}}]{Carraz2012}%
  \BibitemOpen
  \bibfield  {author} {\bibinfo {author} {\bibfnamefont {O.}~\bibnamefont
  {Carraz}}, \bibinfo {author} {\bibfnamefont {R.}~\bibnamefont
  {Charri\`{e}re}}, \bibinfo {author} {\bibfnamefont {M.}~\bibnamefont
  {Cadoret}}, \bibinfo {author} {\bibfnamefont {N.}~\bibnamefont {Zahzam}},
  \bibinfo {author} {\bibfnamefont {Y.}~\bibnamefont {Bidel}}, \ and\ \bibinfo
  {author} {\bibfnamefont {A.}~\bibnamefont {Bresson}},\ }\href@noop {}
  {\bibfield  {journal} {\bibinfo  {journal} {Phys. Rev. A}\ }\textbf {\bibinfo
  {volume} {86}},\ \bibinfo {pages} {033605} (\bibinfo {year}
  {2012})}\BibitemShut {NoStop}%
\bibitem [{\citenamefont {Lautier}\ \emph {et~al.}(2014)\citenamefont
  {Lautier}, \citenamefont {Volodimer}, \citenamefont {Hardin}, \citenamefont
  {Merlet}, \citenamefont {Lours}, \citenamefont {Santos},\ and\ \citenamefont
  {Landragin}}]{Lautier2014}%
  \BibitemOpen
  \bibfield  {author} {\bibinfo {author} {\bibfnamefont {J.}~\bibnamefont
  {Lautier}}, \bibinfo {author} {\bibfnamefont {L.}~\bibnamefont {Volodimer}},
  \bibinfo {author} {\bibfnamefont {T.}~\bibnamefont {Hardin}}, \bibinfo
  {author} {\bibfnamefont {S.}~\bibnamefont {Merlet}}, \bibinfo {author}
  {\bibfnamefont {M.}~\bibnamefont {Lours}}, \bibinfo {author} {\bibfnamefont
  {F.~P.~D.}\ \bibnamefont {Santos}}, \ and\ \bibinfo {author} {\bibfnamefont
  {A.}~\bibnamefont {Landragin}},\ }\href@noop {} {\bibfield  {journal}
  {\bibinfo  {journal} {Appl.Phys.Lett}\ }\textbf {\bibinfo {volume} {105}},\
  \bibinfo {pages} {144102} (\bibinfo {year} {2014})}\BibitemShut {NoStop}%
\bibitem [{\citenamefont {Xu}\ \emph {et~al.}(2017)\citenamefont {Xu},
  \citenamefont {Zhu}, \citenamefont {Zhao}, \citenamefont {Zhang},\ and\
  \citenamefont {Hu}}]{Xu2017}%
  \BibitemOpen
  \bibfield  {author} {\bibinfo {author} {\bibfnamefont {W.-J.}\ \bibnamefont
  {Xu}}, \bibinfo {author} {\bibfnamefont {M.~K.}\ \bibnamefont {Zhu}},
  \bibinfo {author} {\bibfnamefont {M.~M.}\ \bibnamefont {Zhao}}, \bibinfo
  {author} {\bibfnamefont {K.}~\bibnamefont {Zhang}}, \ and\ \bibinfo {author}
  {\bibfnamefont {Z.~K.}\ \bibnamefont {Hu}},\ }\href@noop {} {\bibfield
  {journal} {\bibinfo  {journal} {Phys. Rev. A}\ }\textbf {\bibinfo {volume}
  {96}},\ \bibinfo {pages} {063606} (\bibinfo {year} {2017})}\BibitemShut
  {NoStop}%
\bibitem [{\citenamefont {Dubetsky}\ and\ \citenamefont
  {Kasevich}(2006)}]{Dubetsky2006}%
  \BibitemOpen
  \bibfield  {author} {\bibinfo {author} {\bibfnamefont {B.}~\bibnamefont
  {Dubetsky}}\ and\ \bibinfo {author} {\bibfnamefont {M.~A.}\ \bibnamefont
  {Kasevich}},\ }\href@noop {} {\bibfield  {journal} {\bibinfo  {journal}
  {Phys. Rev. A}\ }\textbf {\bibinfo {volume} {74}},\ \bibinfo {pages} {023615}
  (\bibinfo {year} {2006})}\BibitemShut {NoStop}%
\bibitem [{\citenamefont {Cadoret}\ \emph {et~al.}(2016)\citenamefont
  {Cadoret}, \citenamefont {Zahzam}, \citenamefont {Bidel}, \citenamefont
  {Diboune}, \citenamefont {Bonnin}, \citenamefont {Th\'{e}ron},\ and\
  \citenamefont {Bresson}}]{Cadoret2016}%
  \BibitemOpen
  \bibfield  {author} {\bibinfo {author} {\bibfnamefont {M.}~\bibnamefont
  {Cadoret}}, \bibinfo {author} {\bibfnamefont {N.}~\bibnamefont {Zahzam}},
  \bibinfo {author} {\bibfnamefont {Y.}~\bibnamefont {Bidel}}, \bibinfo
  {author} {\bibfnamefont {C.}~\bibnamefont {Diboune}}, \bibinfo {author}
  {\bibfnamefont {A.}~\bibnamefont {Bonnin}}, \bibinfo {author} {\bibfnamefont
  {F.}~\bibnamefont {Th\'{e}ron}}, \ and\ \bibinfo {author} {\bibfnamefont
  {A.}~\bibnamefont {Bresson}},\ }\href@noop {} {\bibfield  {journal} {\bibinfo
   {journal} {J. Opt. Soc. Am. B}\ }\textbf {\bibinfo {volume} {33}},\ \bibinfo
  {pages} {1777} (\bibinfo {year} {2016})}\BibitemShut {NoStop}%
\bibitem [{\citenamefont {Canuel}\ \emph {et~al.}(2006)\citenamefont {Canuel},
  \citenamefont {Leduc}, \citenamefont {Holleville}, \citenamefont {Gauguet},
  \citenamefont {Fils}, \citenamefont {Virdis}, \citenamefont {Clairon},
  \citenamefont {Dimarcq}, \citenamefont {Bord\'e}, \citenamefont {Landragin},\
  and\ \citenamefont {Bouyer}}]{Canuel2006}%
  \BibitemOpen
  \bibfield  {author} {\bibinfo {author} {\bibfnamefont {B.}~\bibnamefont
  {Canuel}}, \bibinfo {author} {\bibfnamefont {F.}~\bibnamefont {Leduc}},
  \bibinfo {author} {\bibfnamefont {D.}~\bibnamefont {Holleville}}, \bibinfo
  {author} {\bibfnamefont {A.}~\bibnamefont {Gauguet}}, \bibinfo {author}
  {\bibfnamefont {J.}~\bibnamefont {Fils}}, \bibinfo {author} {\bibfnamefont
  {A.}~\bibnamefont {Virdis}}, \bibinfo {author} {\bibfnamefont
  {A.}~\bibnamefont {Clairon}}, \bibinfo {author} {\bibfnamefont
  {N.}~\bibnamefont {Dimarcq}}, \bibinfo {author} {\bibfnamefont {C.~J.}\
  \bibnamefont {Bord\'e}}, \bibinfo {author} {\bibfnamefont {A.}~\bibnamefont
  {Landragin}}, \ and\ \bibinfo {author} {\bibfnamefont {P.}~\bibnamefont
  {Bouyer}},\ }\href@noop {} {\bibfield  {journal} {\bibinfo  {journal} {Phys.
  Rev. Lett.}\ }\textbf {\bibinfo {volume} {97}},\ \bibinfo {pages} {010402}
  (\bibinfo {year} {2006})}\BibitemShut {NoStop}%
\bibitem [{\citenamefont {Wu}\ \emph {et~al.}(2017)\citenamefont {Wu},
  \citenamefont {Zi}, \citenamefont {Dudley}, \citenamefont {Bilotta},
  \citenamefont {Canoza},\ and\ \citenamefont {Muller}}]{Wu2017}%
  \BibitemOpen
  \bibfield  {author} {\bibinfo {author} {\bibfnamefont {X.}~\bibnamefont
  {Wu}}, \bibinfo {author} {\bibfnamefont {F.}~\bibnamefont {Zi}}, \bibinfo
  {author} {\bibfnamefont {J.}~\bibnamefont {Dudley}}, \bibinfo {author}
  {\bibfnamefont {R.~J.}\ \bibnamefont {Bilotta}}, \bibinfo {author}
  {\bibfnamefont {P.}~\bibnamefont {Canoza}}, \ and\ \bibinfo {author}
  {\bibfnamefont {H.}~\bibnamefont {Muller}},\ }\href@noop {} {\bibfield
  {journal} {\bibinfo  {journal} {Optica}\ }\textbf {\bibinfo {volume} {4}},\
  \bibinfo {pages} {1545} (\bibinfo {year} {2017})}\BibitemShut {NoStop}%
\bibitem [{\citenamefont {Geiger}\ \emph {et~al.}(2011)\citenamefont {Geiger},
  \citenamefont {M\'{e}noret}, \citenamefont {Stern}, \citenamefont {Zahzam},
  \citenamefont {Cheinet}, \citenamefont {Battelier}, \citenamefont {Villing},
  \citenamefont {Moron}, \citenamefont {Lours}, \citenamefont {Bidel},
  \citenamefont {Bresson}, \citenamefont {Landragin},\ and\ \citenamefont
  {Bouyer}}]{Geiger2011}%
  \BibitemOpen
  \bibfield  {author} {\bibinfo {author} {\bibfnamefont {R.}~\bibnamefont
  {Geiger}}, \bibinfo {author} {\bibfnamefont {V.}~\bibnamefont {M\'{e}noret}},
  \bibinfo {author} {\bibfnamefont {G.}~\bibnamefont {Stern}}, \bibinfo
  {author} {\bibfnamefont {N.}~\bibnamefont {Zahzam}}, \bibinfo {author}
  {\bibfnamefont {P.}~\bibnamefont {Cheinet}}, \bibinfo {author} {\bibfnamefont
  {B.}~\bibnamefont {Battelier}}, \bibinfo {author} {\bibfnamefont
  {A.}~\bibnamefont {Villing}}, \bibinfo {author} {\bibfnamefont
  {F.}~\bibnamefont {Moron}}, \bibinfo {author} {\bibfnamefont
  {M.}~\bibnamefont {Lours}}, \bibinfo {author} {\bibfnamefont
  {Y.}~\bibnamefont {Bidel}}, \bibinfo {author} {\bibfnamefont
  {A.}~\bibnamefont {Bresson}}, \bibinfo {author} {\bibfnamefont
  {A.}~\bibnamefont {Landragin}}, \ and\ \bibinfo {author} {\bibfnamefont
  {P.}~\bibnamefont {Bouyer}},\ }\href@noop {} {\bibfield  {journal} {\bibinfo
  {journal} {Nat. Commun.}\ }\textbf {\bibinfo {volume} {2}},\ \bibinfo {pages}
  {474} (\bibinfo {year} {2011})}\BibitemShut {NoStop}%
\bibitem [{\citenamefont {Barrett}\ \emph {et~al.}(2016)\citenamefont
  {Barrett}, \citenamefont {Antoni-Micollier}, \citenamefont {Chichet},
  \citenamefont {Battelier}, \citenamefont {L\'{e}v\`{e}que}, \citenamefont
  {Landragin},\ and\ \citenamefont {Bouyer}}]{Barrett2016}%
  \BibitemOpen
  \bibfield  {author} {\bibinfo {author} {\bibfnamefont {B.}~\bibnamefont
  {Barrett}}, \bibinfo {author} {\bibfnamefont {L.}~\bibnamefont
  {Antoni-Micollier}}, \bibinfo {author} {\bibfnamefont {L.}~\bibnamefont
  {Chichet}}, \bibinfo {author} {\bibfnamefont {B.}~\bibnamefont {Battelier}},
  \bibinfo {author} {\bibfnamefont {T.}~\bibnamefont {L\'{e}v\`{e}que}},
  \bibinfo {author} {\bibfnamefont {A.}~\bibnamefont {Landragin}}, \ and\
  \bibinfo {author} {\bibfnamefont {P.}~\bibnamefont {Bouyer}},\ }\href@noop {}
  {\bibfield  {journal} {\bibinfo  {journal} {Nature Commun.}\ }\textbf
  {\bibinfo {volume} {7}},\ \bibinfo {pages} {13786} (\bibinfo {year}
  {2016})}\BibitemShut {NoStop}%
\end{thebibliography}%

\end{document}